\documentclass[useAMS,usenatbib]{mn2e}
\usepackage{graphicx}
\usepackage{amsmath,epsfig,rotating,amssymb}
\usepackage{listings}
\usepackage{amsmath}
\title[Outflows and mass accretion in dense cores]{Outflows and mass accretion in collapsing dense cores with misaligned rotation axis and magnetic field}
\author[Andrea Ciardi and Patrick Hennebelle]{Andrea Ciardi\thanks{E-mail:
andrea.ciardi@obspm.fr} and Patrick Hennebelle\\
LERMA, Obervatoire de Paris, {\'E}cole Normale Sup{\'e}rieure, Universit{\'e} Pierre et Marie Curie, UMR 8112 CNRS\\24 Rue Lhomond, 75231 Paris, France\\}

\begin{document}

\date{Accepted . Received ; in original form}

\pagerange{\pageref{firstpage}--\pageref{lastpage}} \pubyear{2010}

\maketitle

\label{firstpage}

\begin{abstract}
Outflows and jets are intimately related to the formation of stars, and play an important role in redistributing mass, energy and
angular momentum within the dense core and parent cloud. The interplay
between magnetic field and rotation is responsible for launching these
outflows, whose formation has been generally carried out for idealized
systems where the angle $\alpha$ between the rotation axis and large-scale magnetic
field is zero. Here we explore, through three-dimensional ideal magneto-hydrodynamic
simulations, the effects of a non-zero $\alpha$ on the formation of
outflows during the collapse of dense pre-stellar cores. We find that mass ejection is less efficient for increasing angle
$\alpha$, and that outflows are essentially suppressed for $\alpha\sim90^{\circ}$. An
important consequence is a corresponding increase of the mass accreted
onto the adiabatic (first) core. In addition, mean flow velocities tend to increase with $\alpha$, and misaligned configurations
produce clumpy, heterogeneous outflows that undergo precession, and are more prone to instabilities.
\end{abstract}


\section{Introduction}

The ejection of plasma in the form of more or less collimated bi-polar
flows (jets) is arguably one of the most spectacular displays staged
by a forming star. Over a million years, these outflows trace the
creation of stars from their embryonic emergence as proto-stars, still
enveloped in their dense parental cloud, to their birth
as T Tauri stars surrounded by proto-planetary discs. The existence
of a close relation between the ejection and accretion of matter throughout
the (low-mass $M_{\star}\lesssim2M_{\odot}$) star's formation history
\citep{cabrit_co_1992,hartigan_disc_1995,wu_study_2004} suggests
the presence of a single launching mechanism, which is now widely
thought to rest on the presence of a large-scale magnetic field mediating
the extraction of gravitational energy from the accreting plasma,
and redirecting mass and energy, in the form of kinetic and Poynting
flux, into bi-polar jets. The fundamental magneto-hydrodynamic (MHD)
mechanism for launching and collimating an axisymmetric, self-similar
outflow was detailed in analytical work by \cite{blandford_hydromagnetic_1982}
and applied to young stellar systems by \cite{pudritz_centrifugally_1983};
it was later extended to include non-self similar solutions \cite{pelletier_hydromagnetic_1992}
and global disc-jet solutions by \cite{ferreira_magnetically-driven_1997}.
The basic ideas underlying MHD jet launching were confirmed in time-dependent
two-dimensional, axisymmetric numerical simulations \citep{uchida_1985} which either
included a disc \citep{kudoh_magnetically_1998,zanni_mhd_2007} or
prescribed it as boundary conditions \citep{ouyed_numerical_1997,ustyugova_magnetocentrifugally_1999,anderson_structure_2005,
fendt_collimation_2006,pudritz_controllingcollimation_2006}.
Furthermore, the early ejection of bipolar outflows was also demonstrated in two- \citep{mellon_magnetic_2008} and
three-dimensional MHD simulations of the collapse of pre-stellar
dense cores \citep{hennebelle_magnetic_2008}, which notably included
the second collapse and the formation of jets from the protostar \citep{machida_first_2006,banerjee_outflows_2006,machida_high-_2008}.
In fact, far from being a passive tracers of star formation, these
outflows play an active role which is still to be fully understood.
Examples include the removal of angular momentum from the accreting
flow \citep{bacciotti_hubble_2002}, determining the efficiency of
star formation by dispersing the accreting envelope \citep{matzner_efficiencies_2000},
and sustaining turbulence in their local environment \citep{nakamura_protostellar_2007}.

In general, the modelling of collapsing cores and jet launching has been carried out under
the simplifying assumption that the angle $\alpha$ between the rotation
axis and the large-scale magnetic field, is zero. Exceptions are the
simulations of \cite{machida_evolution_2006,price_impact_2007,hennebelle_disc_2009},
which however focused mostly on the core and disc dynamics. Here we present simulations of the collapse of pre-stellar
dense cores with misaligned initial configurations ($\alpha=0^{\circ}-90^{\circ}$),
and show that the angle $\alpha$ is fundamental in determining not
only the properties of the outflow, but also the mass accretion onto
the core. In particular we observe the gradual decrease in the mass
ejection with increasing angle $\alpha$, until its total suppression
for nearly perpendicular configurations.

\section{Results}
\subsection{The numerical model}
We follow numerically the gravitational collapse of dense pre-stellar
cores, up to, and including the formation of the first (or adiabatic)
core. The numerical simulations were performed with Ramses \citep{teyssier_cosmological_2002,fromang_high_2006},
an adaptive mesh refinement code which uses a Godunov-type scheme
and constrained transport to solve the ideal MHD
equations. Throughout the simulations the Jeans length is resolved
with at least 10 cells, and a HLLD solver is employed. The initial
conditions consist of a one solar mass core whose density profile
resembles the observed cores, and is given by $n(r)=n_{c}/\left[1+(r/r_{0})^{2}\right]$
where $r_{0}\sim1000$ AU is the inner cloud radius. This isothermal
cloud ($T\sim10$ K) is placed inside a warm and diffuse medium in
pressure equilibrium with the cloud edge, with a contrast of 10 between
central, $n_{c}=8\times10^{6}$ cm$^{-3}$, and edge densities. The
cloud is initially in solid body rotation and threaded by a uniform
magnetic field along the $z$-axis, whose intensity is proportional
to the column density of the cloud. The rotation axis is in the $x-z$
plane and makes an angle $\alpha$ with respect to the magnetic field.
For ease of presentation we define, in addition to the Cartesian coordinates
$(x,y,z)$, cylindrical coordinates $(\varpi,\phi,Z)$ with $Z$ parallel
to the initial rotation axis. All simulations are characterized by
four parameters: the ratio of rotational over gravitational energy
($\simeq0.03$), the ratio of thermal over gravitational energy ($\simeq0.25$),
the degree of magnetization $\mu$, and the angle $\alpha$. Only
the effects of different $\mu$, the mass-to-flux 
over critical mass-to-flux ratio, and $\alpha$ are investigated in
this work.

\begin{figure*}
\includegraphics*[width=175mm]{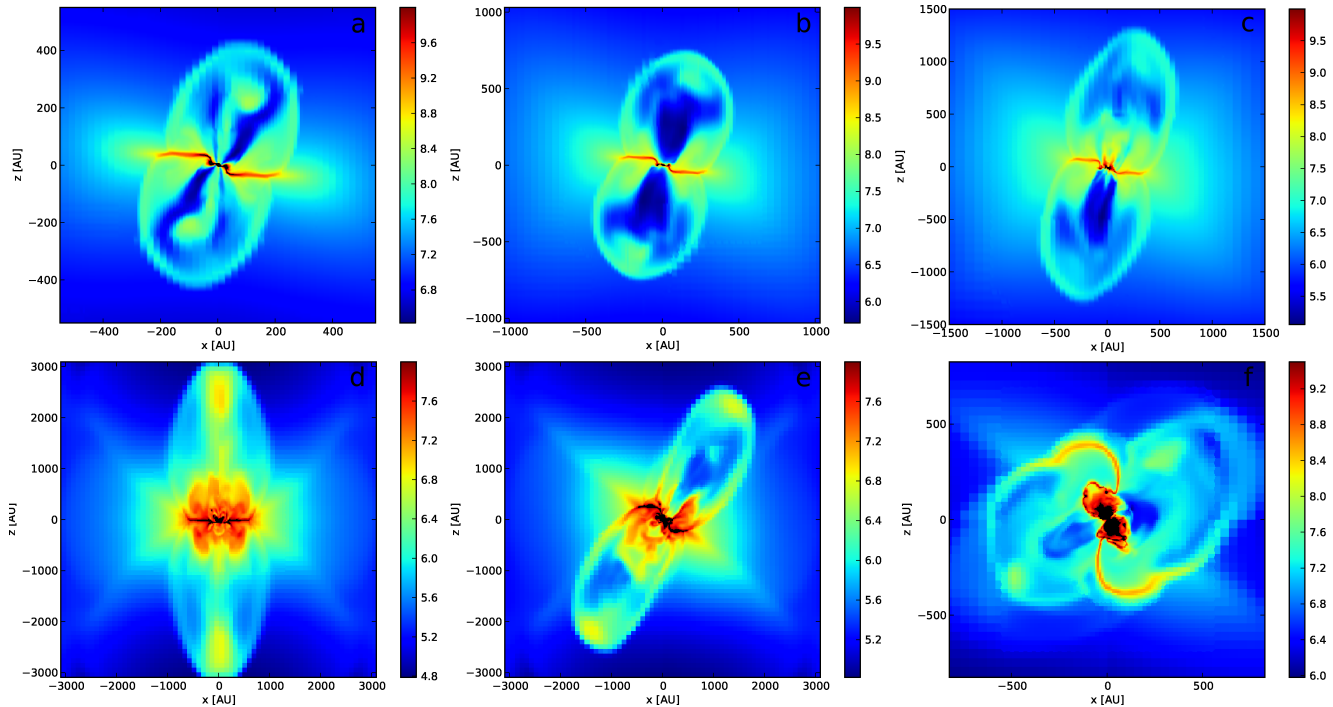}
\caption{Slices in $x-z$ plane of the particle number density (cm$^{-3}$ on a logarithmic scale) for the case $\mu=5$. The first row shows for $\alpha = 20^\circ$ the times: (a) 20300 years, (b) 21300 years, and (c) 22200 years. The second row shows at $\sim 27000$ years the angles: (d) $\alpha = 0^\circ$, (e) $\alpha = 45^\circ$ and (e) $\alpha = 70^\circ$. Note the use of the different scales between images.}
\label{figure1}
\end{figure*}

\subsection{Outflow dynamics, precession and instabilities}
\begin{figure}
 \includegraphics[width=85mm]{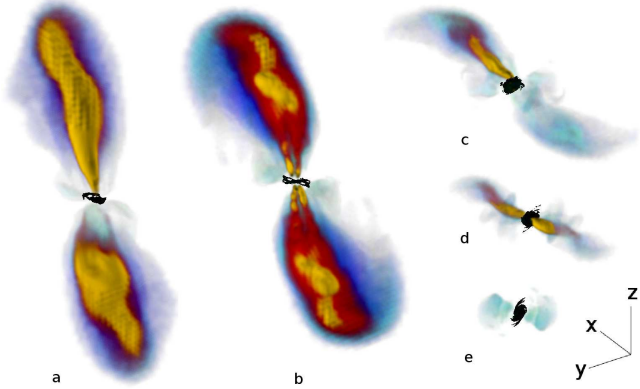}
 \caption{Volume rendering of the outflows at $\sim23000$ years, for $\mu=5$ and $\alpha$ equal to (a) $20^\circ$, (b) $45^\circ$, (c) $70^\circ$, (d) $80^\circ$ and (e) $90^\circ$. The colours show increasing speeds (from blue to yellow) in an onion-like structure. The black disc-like regions represent densities $>10^{10}$ cm$^{-3}$.}
 \label{figure2}
\end{figure}

The general outflow formation and dynamics are presented in Figure~\ref{figure1}. As the initially spherical, magnetized pre-stellar core undergoes
gravitational collapse it flattens preferentially along the magnetic
field lines, producing a dynamically-collapsing, magnetized disc-like
structure or \emph{pseudo-disc}. Following the isothermal collapse
phase, an adiabatic core with densities $\gtrsim10^{10}$ cm{\scriptsize $^{-3}$}
and a radius $\sim10-20$ AU, forms at the centre of the infalling envelope.
The ensuing build-up of a centrifugally supported disc (or simply
\textit{disc}), with characteristic diameters $\sim50-200$ AU, depends
on the efficiency of magnetic torques to remove angular momentum from
the pseudo-disc. In \cite{hennebelle_disc_2009} it was argued that
the magnetic braking efficiency is proportional to $h^{-1/2}$, where
the characteristic scale-height $h$ of the disc is in turn determined by the angle
$\alpha$ between the initial magnetic field and rotation axis. Increasing
their misalignment produces thicker pseudo-discs which are less efficiently
braked by the magnetic field, and thus lead more easily to the formation
of centrifugally supported discs. In general we find that outflows are launched for all angles $\alpha\lesssim80^{\circ}$
independently of the existence of a centrifugally supported disc.
However when $\alpha\sim90^{\circ}$ the ejection of matter
is essentially suppressed, even when a disc is present. The remaining outward motions
observed in the perpendicular configuration are particularly interesting because mass ejection does not produce either a magnetic cavity, or
a jet. In fact, mass is not launched from the core, but the \emph{outflow} traces dense regions
of the disc which are at a large radii from the core ($\sim200$ AU)
and are being radially expelled by the highly twisted magnetic field.
The outflowing gas propagates perpendicularly to the rotation axis (see
Figure ~\ref{figure2}) and in the plane of the disc, with radial
ejection and azimuthal speeds, $v_{\varpi}\approx v_{\phi}$.

For $\alpha\lesssim80^{\circ}$ the formation of bipolar outflow begins
as the magnetic field lines, connecting the outer regions of the cloud
to its rapidly rotating inner regions, undergo strong shear which
results in a significant azimuthal component ($B_{\phi}$) of the
magnetic field being generated close to the adiabatic core and disc
(if present). The growing magnetic pressure gradient accelerates the
plasma, inflating bi-polar magnetic cavities within the infalling
envelope. At this stage the ratio of MHD Poynting to kinetic flux magnitudes, $\Gamma = v_{\perp}B^{2}/2\pi\rho v^{3}$, where  $v_{\perp}$ is the component of the velocity perpendicular to the magnetic field, and the
plasma-$\beta$, defined as the ratio of the thermal to magnetic pressure,
are typically $\Gamma \sim 2-10$, and $\beta\sim10^{-3} - 0.1$ in the magnetic cavity.  Such ``magnetic
tower'' structure is similar to that described by \cite{lynden-bell_discs_2003} and re-produced in scaled laboratory experiments by \cite{lebedev_magnetic_2005}, and it is inflated by the magnetic field and is not mechanically-driven by wide-angle winds.  At
the base of the cavity the continuous generation of $B_{\phi}$ provides
the Poynting flux powering an outflow which originates from the core
and a region extending several AU around it. Depending on the ejection
efficiency, which is measured by the ejection index  $\xi=d\ln\dot{M}_{a}/d\ln\varpi$ with $\dot{M_{a}}$
the mass accretion rate through the disc, this outflow can be described
as either magneto-centrifugally or magnetic pressure driven \citep{ferreira_magnetized_1995}.
As we shall see in the next section, the ejection efficiency increase
in time and both launching regimes are attained during the simulations.
The collimation of this ``disc'' wind into a jet depends on the
radial distribution of currents (e.g. \citealt{pudritz_controllingcollimation_2006}),
and therefore on the width of the magnetic cavity itself, which takes
up the ``return current'' and the magnetic stresses associated
with the $B_{\phi}$ component of the magnetic field (e.g. \citealt{spruit_theory_2009}).
However, the collimation of the magnetic tower depends in
turn on the distribution of gas and magnetic field in the surrounding
environment, which is swept up into a shock layer delineating
the walls of the expanding magnetic tower. Therefore the medium through which the magnetic tower expands plays as well an important role on the overall collimation of the outflow. Figure \ref{figure2} shows the outflows' three-dimensional structure for different $\alpha$, indicating that increasing the misalignment leads to the narrowing
and better collimation of the whole magnetic tower. This is due to the increasing scale-height of the pseudo-disc with $\alpha$, which provides confinement over a longer lenght of the magnetic tower. In addition, because the outflow is inclined with respect to the pseudo-disc, it tends to propagates into it, leading again to a better lateral confinement, albeit asymmetric. The difference between the aligned and misaligned cases is also evident in Figure \ref{figure1}d and e, where the base of the magnetic cavity, which is still embedded in the pseudo-disc, is indeed narrower for the misaligned case.

The angle $\alpha$ has also a strong impact on the homogeneity of
the outflows. We find that for misaligned configurations the outflows undergo precession around the $z$-axis, and that this effect is smaller for larger values of $\alpha$, settling within $\sim2000$ years to an approximately constant angle $\theta$, which we remind is initially
zero. Nutation is very small, and the initial angle between the magnetic field and rotation axis, $\alpha$, remains approximately constant. Typical values of $(\alpha,\theta)$, for $\mu=5$, are $(20^{\circ},60^{\circ})$,
$(45^{\circ},25^{\circ})$, $(70^{\circ},20^{\circ})$ and $(80^{\circ},10^{\circ})$. In addition, while the magnetic tower is initially launched perpendicular
to the pseudo-disc, thus approximately parallel to the magnetic field, the outflow is ejected parallel to the rotation axis of the precessing core/disc, resulting in asymmetric cavities as shown in Figure~\ref{figure1} and \ref{figure2}. At later times the spatial variability observed  in the outflows is mainly the consequence of growing non-axisymmetric current-driven modes and episodic ejections. The unsteady behaviour appears to be related to the increasing mass ejection efficiency, and build up of $B_{\phi}$ in the core/disc \citep{anderson_structure_2005}. The precession present in the misaligned models further helps to seed large kink-like perturbations in the jet body, thus promoting the growth of the instability, which is otherwise artificially suppressed in the aligned case. The presence of a kink-like instability is indeed expected in jets confined by a helical magnetic field, and in detailed three-dimensional simulations it was also seen to lead to considerable distortions of the jet body and its fragmentation \citep{moll_kink_2008}. In our simulations we find that the outflows are above the Kruskal-Shafranov stability threshold, with typical values of the magnetic pitch evaluated along single field lines, $ \varpi|B_Z|/ |B_{\phi}| \lesssim 30$ AU (e.g. \citealt{appl_current-driven_2000}). Precession and instabilities therefore lead to the fragmentation of the outflow into clumps, which together with the generation of episodic ejection, produce a series of nested cavities and internal shocks, such as those visible in Figure \ref{figure1}e and f, where faster ejections plough through previously launched material. Scaled laboratory astrophysics experiments of compressible, super-magnetosonic MHD flows have also shown that although jets may be episodic and unstable, multiple ejections of magnetic cavities and jets can produce clumpy flows which remain well collimated \citep{ciardi_episodic_2009}.

\subsection{Outflow suppression and mass accretion.}
\begin{figure}
 \includegraphics[width=85mm]{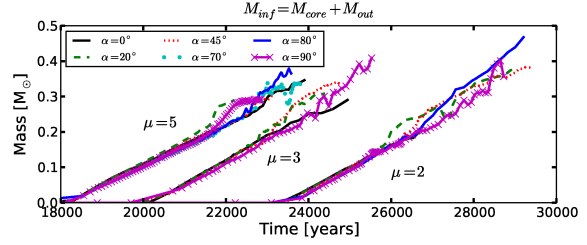}
 \caption{Total mass $M_{core}+M_{out}$ for different magnetizations $\mu$ and angles $\alpha$.}
 \label{figure3}
\end{figure}

\begin{figure}
\includegraphics[width=80mm]{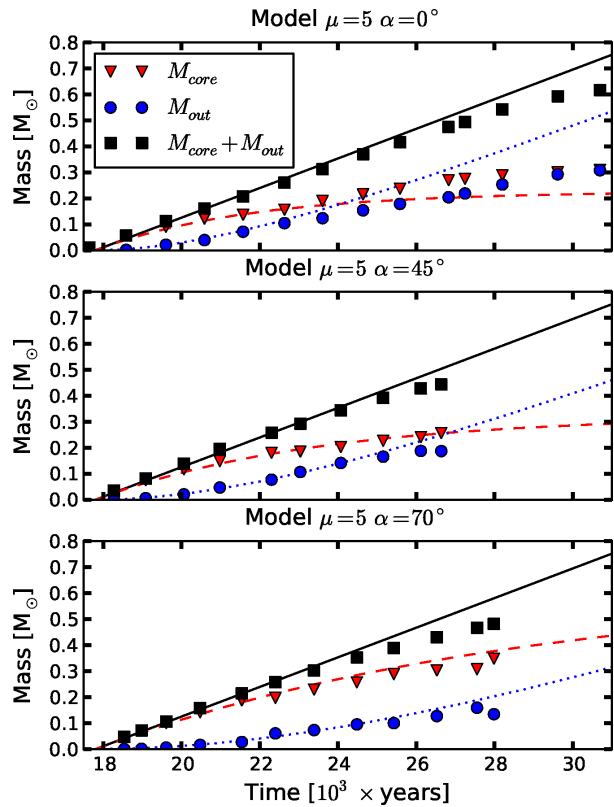}
\caption{Total mass in the core (dashed) and outflow (dotted) for the analytical model, calculated using the nominal value of the mass infall (solid line) rate taken from \protect\cite{hunter_collapse_1977}, $\dot{M}_{inf}=36c_{s}^{3}/G=5.7\times10^{-5}$ M$_{\odot}$ yr$^{-1}$, and the accretion-ejection time-scale from the aligned case $\tau_{0}=4000$ years. The symbols are data from the simulations.}
\label{figure4}
\end{figure}

We find that the general trend of mass ejection-accretion during the
collapse of a pre-stellar core, is a decrease in the mass ejection
efficiency for increasing $\alpha$, which leads to a corresponding increase in the total mass, $M_{core}(t)$, accreted
onto the central core. Figure~\ref{figure3} shows the total mass
in the core and outflow ($M_{out}+M_{core}$) as a
function of time. The outflow is defined as the plasma with $\mathbf{v\cdot}\mathbf{\hat{R}}>0.1$
km s$^{-1}$, where $\mathbf{\hat{R}}$ is the unit spherical radius; the core is defined as the plasma with a particle number density $n>10^{9}$
cm$^{-3}$ and $\mathbf{v\cdot}\mathbf{\hat{R}}\leq0.1$ km s$^{-1}$.
From mass conservation the sum $M_{out}+M_{core}$ is the total \emph{infall
mass}, $M_{inf}$, which at early times is seen to be approximately constant,
and largely independent of both angle $\alpha$ and magnetization
$\mu$. Its value is consistent with that given by \cite{hunter_collapse_1977},
who considered similarity solutions of the collapse of an unstable
isothermal sphere. The relation between the accreted-ejected mass, and the misalignment
of rotation axis and magnetic field, can be understood by considering
the total mass outflow rate, $\int d\dot{M}_{out}=\int\rho\mathbf{v}\cdot d\mathbf{S}$.
For steady-state, axisymmetric flows it may be written as $\int d\dot{M}_{out}=\int\kappa d\Phi=\int\kappa\mathbf{B_{p}\cdot}d\mathbf{S}$,
where $\kappa=4\pi\rho v_{P}/B_{P}$ is the mass load per unit time, per
unit poloidal magnetic flux, and may also be written in terms of the density, $\rho_A$, at the Alfv\'en surface as $\kappa=\sqrt{\rho_A}$ (\citealt{pelletier_hydromagnetic_1992}).
We make the simplifying assumptions that $\kappa$ and $\mathbf{B_{P}\cdot}\mathbf{\hat{S}}\sim B_{z}\cos\alpha$
are constant over the surface ($S$) of the core, so that the integrated
mass outflow rate can be approximated by $\dot{M}_{out}\sim\kappa\Phi\sim\kappa B_{z}\cos\alpha\, S$.
We now proceed to establish a scaling relation between the
mass ejection rate, $\dot{M}_{out}$, and the mass in the core, $M_{core}$.
We assume that the core, of characteristic radius $R$, is in hydrostatic
equilibrium, $GM_{core}/R\propto PR^{-3}$, and threaded by a constant
magnetic flux, $\phi\propto BR^{2}\sim\text{const}$. The pressure
$P$ and density $\rho$ scale as $P\propto\rho^{\gamma}$ and $\rho\propto M_{core}R^{-3}$
respectively, and combining these expression one finds: $R\propto M_{core}^{(2-\gamma)/(4-3\gamma)}$,
$\rho\propto M_{core}^{-2/(4-3\gamma)}$, $S\propto R^{2}\propto M_{core}^{(4-2\gamma)/(4-3\gamma)}$,
$B\propto M_{core}^{(2\gamma-4)/(4-3\gamma)}$, and $\kappa\propto \rho^{1/2} \propto M_{core}^{-1/(4-3\gamma)}$. The relation obtained for $\kappa$ is equivalent to assuming that the launching speed at the core/disc, $v$, scales with the Alfv\'en speed, $B/\sqrt{4\pi\rho}$. From these relations, the mass outflow rate is found to scale as a power of the total mass present
in the core, $\dot{M}_{out}\propto M_{core}^{-1/(4-3\gamma)}$, which
for the adiabatic core ($\gamma=5/3)$ gives the linear relation:
\begin{equation}
\dot{M}_{out}=\eta M_{core}(t)\label{eq:1}\end{equation}
The characteristic accretion-ejection time-scale,
$\eta^{-1}=\tau_{ae}=\tau_{0}/\cos\alpha=const$, depends only
on the angle $\alpha$, where $\tau_{0}$ is the value corresponding
to the aligned case ($\alpha=0^{\circ}$) which is to be determined numerically.
We can then combine Eq. \ref{eq:1} with the expression for the
mass conservation, $\dot{M}_{inf}=\dot{M}_{core}+\dot{M}_{out}$,
to solve for the mass accreted onto the core as a function of time:
\begin{equation}
M_{core}(t)=\tau_{ae}\dot{M}_{inf}(1-e^{-t/\tau_{ae}})\label{eq:core}\end{equation}
and for the total ejected mass: \begin{equation}
M_{out}(t)=\tau_{ae}\dot{M}_{inf}(e^{-t/\tau_{ae}}-1)+\dot{M}_{inf}\, t\label{eq:out}\end{equation}
Figure~\ref{figure4} shows the simulated infall, core and outflow masses for
the case $\mu=5$, and for different angles $\alpha$, together with
the analytical solutions, which are strictly valid in the regime where $\dot{M}_{inf}=const$. The results show that at early times the mass in the core is small, the ejection rate is negligible,
and mass accretion dominates, $\dot{M}_{core}=\dot{M}_{inf}e^{-t/\tau_{ae}}$, over the mass ejection, which is given by $\dot{M}_{out}=\dot{M}_{inf}(1-e^{-t/\tau_{ae}})$. The model predicts that within a time-scale
$\tau_{ae}=\tau_{0}/\cos\alpha$ the core accretes $\sim63\%$ of
its ``final'' mass, which is asymptotically given by $M_{C}=\tau_{ae}\dot{M}_{inf}$ (for $\alpha\neq90^{\circ}$). Since $\tau_{ae}$ increases with the angle $\alpha$, the initial
misalignment between the rotation axis and magnetic field plays a
fundamental role in determining the masses of the cores. The limiting
case is for a perpendicular magnetic field and rotation axis ($\alpha=90^{\circ}$),
where no mass is ejected from the core, and its growth is at the (time-dependent) mass
infall rate $\dot{M}_{inf}$. This case being qualitatively similar
to hydrodynamical simulations where no outflows are produced.
The high ejection efficiency predicted by the model at late times is however not
energetically favourable. A fraction of the outflow fails to become
trans-Alfv\'enic and eventually falls back onto the core/disc before
being reprocessed. By combining the mass conservation equation with the definition of $\xi$, the time when the limiting ejection efficiency occurs ($\xi=1$ for jets with negligible enthalpy) can be estimated from the relation: $\left(\varpi_{i}/\varpi_{e}\right)^{\xi}=1-\dot{M}_{out}/\dot{M}_{inf}$ \citep{ferreira_magnetized_1995}; where $\varpi_{i}\sim10-20$ AU corresponds to the core radius, and $\varpi_{e}\sim50-100$ AU corresponds to the outer radius of the accretion-ejection structure, where $\dot{M}_{inf}$ is evaluated. For our model, $\xi=t\left[\ln\left(\varpi_{e}/\varpi_{i}\right)\tau_{ae}\right]^{-1}$, and the time when we expect the the analytical model to overestimate the ejected mass is when $\xi\sim1$, which gives $t\sim(1-3)\times\tau_{ae}$. This occurs for the aligned case, which is the only model to have reached a high-ejection efficiency, at $\sim24000$ years, and it is consistent with our estimates. From inspection of the velocity fields, it is also evident that part of the outflow is failing to be ejected and falls back as a lower density envelope ($n<10^{10}$ cm $^{-3}$). This material is visible in the first panel of Figure~\ref{figure4} as an increase in the core's mass with respect to the analytical solution.

The results show that with increasing misalignment, the general trend of mass ejection-accretion in collapsing pre-stellar cores is to produce
higher mass adiabatic cores, with lower ejection efficiency ($\xi\propto\cos\alpha$).
Another consequence is an increase in the bulk speed of the outflow,
defined as $v_{b}=\int dV\rho|\mathbf{v}|/\int dV\rho$, with the
integral taken over the volume of plasma where $\mathbf{v\cdot}\mathbf{\hat{R}}>0.1$
km s$^{-1}$. Typical values at $\sim26000$ years are $v_{b}(\alpha=0^{\circ})=1.17$
km s$^{-1}$, $v_{b}(\alpha=45^{\circ})=1.43$ km s$^{-1}$ and $v_{b}(\alpha=70^{\circ})=1.76$
km s$^{-1}$, with their ratios consistent with the angle dependence
obtained in the analytical model. This increase in the bulk speed can be understood by considering the expression for the
asymptotic jet velocity $u_{\infty}=\Omega\varpi\sqrt{(2\lambda-3)}$
(\cite{blandford_hydromagnetic_1982}). For misaligned, collapsing pre-stellar cores two
factors contribute to the higher bulk speeds. The first is related to the magnetic
lever arm $\lambda$, which depends on the ejection index through the relation
$\lambda\simeq1+\frac{1}{2\xi}$ \citep{ferreira_magnetically-driven_1997};
the second is related to the rotational speed, which scales with the core's mass, $\Omega\varpi\sim\sqrt{GM_{core}/R}$.
Thus at a given time, the more massive
cores obtained for larger $\alpha$, and the corresponding lower ejection efficiency, combine to eject faster outflows.
Finally we note that following the initial launching of the magnetic tower, as $\xi$ increases in time the outflow from the core/disc is
first launched with a relatively large magnetic lever arm, and
may be thought of as magneto-centrifugally driven; at later times, with
increasing $\xi$, the lever arm is small and the outflow may be better described as magnetic
pressure driven.

\section{Summary}

We have presented three-dimensional simulations of the collapse of
pre-stellar dense cores with misaligned magnetic field and rotation
axis. Imposing non-axisymmetric initial conditions has several, important
effects on the formation of the first adiabatic core and its associated
outflows. In particular an increasing angle $\alpha$
produces a decrease in the ejection efficiency, leading to the suppression
of outflows for $\alpha\sim90^{\circ},$ and a corresponding increase
of the mass of the cores. In addition we find that the misalignment of magnetic field and rotation
axis produces precessing jets which are more susceptible to current-driven
instabilities, leading to the formation of heterogeneous, clumpy
flows. These effects are visible even for small $\alpha$, and indicate that aligned configurations may be too idealized, and unable to capture the whole dynamics. To interpret the numerical results we have developed a model which seems to capture the essential features observed in the simulations. In particular we find that for low ejection efficiencies the mass ejection rate, $\dot{M}_{out}$, is proportional to the mass of the core, and that the reduction of $\dot{M}_{out}$ with $\alpha$ is consistent with a decrease of the magnetic flux threading the (approximately constant) core's surface.

\section*{Acknowledgments}
We acknowledge the support of the CINES and CEMAG computing centers. Part of the work was supported by the Marie Curie Reintegration Grant, MAGPLUS.

\end{document}